\documentstyle[aps,graphicx,12pt]{revtex}

\def\[{\left[}
\def\]{\right]}
\def\nn{\nonumber}
\def\({\left(}
\def\){\right)}
\def\labels#1{\label{#1}}
\def\eq#1{(\ref{#1})}
\def\d{\delta}
\def\s{\sigma}

\def\.{\cdot}

\def\.{\!\cdot\!}
\def\bi{\begin{itemize}}

\def\ei{\end{itemize}}
\def\be{\begin{eqnarray}}
\def\ee{\end{eqnarray}}
\def\bn{\begin{enumerate}}
\def\en{\end{enumerate}}
\def\h{{1\over 2}}
\def\n{\noindent}
\def\nn{\nonumber}

\def\bk#1{\langle#1\rangle}
\def\r2{\sqrt{2}}

\def\x{\times}

\def\eq#1{(\ref{#1})}

\def\o{\omega}

\def\s{\sigma}

\def\d{\ \delta}

\def\rr2{{1\over\sqrt{2}}}

\def\b#1{\tilde{#1}}

\def\Tr{{\rm Tr}}
\def\O{\Omega}

\begin{document}

\title{Invariant Integration over the Unitary Group}
\author{S. Aubert and C.S. Lam}
\address{Department of Physics, McGill University\\
3600 University St., Montreal, Q.C., Canada H3A 2T8\\
Email: samuel.aubert@elf.mcgill.ca, Lam@physics.mcgill.ca}

\maketitle

\begin{center}
{\bf Abstract}
\end{center}
\begin{abstract}
Integrals for the product of unitary-matrix elements 
over the $U(n)$ group will be discussed.
A group-theoretical formula is available to convert them
 into a multiple sum, but unfortunately the sums are 
often tedious to compute. In this paper, we develop an alternative method 
in which these sums are avoided, and group theory is rendered unnecessary.
Only unitarity and the invariance of
the Haar measure are required for the computation. 
The method can also be used to get a closed expression for the
simpler integral of monomials over a hypersphere.
\end{abstract}

\section{Introduction}
The integral \be
\int (dU)\ U^*_{i_1j_1}\cdots U^*_{i_pj_p}
U_{k_1l_1}\cdots U_{k_pl_p}\nn\ee
for the product of $n\x n$ unitary matrix elements
and their generating functions are useful in many
areas of physics. That includes two-dimensional
quantum gravity \cite{DGZ-J}, QCD, matrix models, and
statistical and condensed-matter problems of various
sorts \cite{GM-GW}. These integrals are also useful in
the parton saturation problem at small Feynman-$x$ \cite{LMZ}.
The monomial integral above can be computed using a graphical
method \cite{Cr}. A more powerful expression can be
derived using
the Itzykson-Zuber formula \cite{IZ} as a generating function,
or directly from group theory \cite{GT}
using the Frobenius formula \cite{Weyl}.
Simplification can be obtained for $n\to\infty$ \cite{LN}, but for finite $n$,
the expression is quite complicated (see eq.~\eq{gpint}).
It involves multiple sums over an expression containing
characters of the symmetric group $S_p$, as well as the dimensions
of irreducible representations of $S_p$ and of the unitary group
$U(n)$. One of the sums is taken over all the relevant irreducible
representations, and the others are taken over the symmetry
groups of the index sets $I=(i_1\cdots i_p)$ and $J=(j_1\cdots j_p)$.
These sums could be long and tedious for large $p$, and for indices
which have a high degree of symmetry.

In this paper, we discuss an alternative method to calculate the
monomial integral, using as input only the unitarity of the matrices
$U$, and the invariance of the Haar measure $dU$. No knowledge
of group theory is required, and these complicated multiple
sums are avoided. We shall refer to this 
method as the `invariant method'.

Invariance of the group measure produces various relations
between the different integrals, which will be discussed in 
Sec.~III A. The values of these integrals are obtained recursively
from the unitarity relation, and that is discussed in Secs.~III B to  III D.  

The invariant method is also applicable to the much simpler case
of a monomial integral over a hypersphere.
This simpler case will be used as a testing ground for the idea.
It will be discussed in Sec.~II, as a preparation
for the computation of the unitary integral in Sec.~III.

In a forthcoming paper, the relative advantages of the invariant method
and the group-theretical
formula will be discussed. We will also examine relations that can be
obtained by combining both approaches.

\section{Integration over a Hypersphere}
Let $\O_{n-1}$ be the unit sphere in $n$ dimensions, defined by
\be
\sum_{i=1}^n x_i^2=1,\labels{unitsphere}\ee
and $(d\o)$ be its rotationally symmetric volume element,
normalized to $\int (d\o)=1$. 
We wish to calculate the integral $\int(d\o)Y$ over $\O_{n-1}$, where $Y$
is a monomial of the coordinates $x_j\ (j=1,2,\cdots,n)$.
This integral is zero unless the power of every $x_j$ is even,
 in which case it can be written in
the form
\be
\bk{J|J}=\int (d\o)\ X_JX_J,\labels{sphereint}\ee
where $X_J\equiv \prod_{a=1}^{p}x_{j_a}$ is a monomial of  degree
$p$, indexed by the set $J=(j_1j_2\cdots j_p)$.

One can attempt to calculate the integral in several ways. Three of them
having analogs with the $U(n)$ integrals will be singled out,
because the simpler setting of a sphere should make their relative merits
more transparent.
The first two are standard, both using
 the spherical coordinates to calculate.
The third one, which we wish to develop in this paper,
is an {\it invariant approach}, requiring no coordinate
system in its computation.

\subsection{Spherical coordinates}
The spherical coordinates of $x_j$ on the unit sphere are:
\be
x_n&=&cos\theta_1\nn\\
x_{n-1}&=&\sin\theta_1\cos\theta_2\nn\\
x_{n-2}&=&\sin\theta_1\sin\theta_2\cos\theta_3\nn\\
\cdots&&\nn\\
x_2&=&\(\prod_{i=1}^{n-2}\sin\theta_i\)\cos\phi\nn\\
x_1&=&\(\prod_{i=1}^{n-2}\sin\theta_i\)\sin\phi.\labels{sc}\ee
The range of $\theta_i$ is between 0 and $\pi$, and the range of $\phi$
is between 0 and $2\pi$.
The volume element is 
\be
(d\o')&=&\(\prod_{i=1}^{n-2}(\sin\theta_i)^{n-i-1}d\theta_i\)d\phi\nn\\
(d\o)&=&{(d\o')\over\int (d\o')}.\labels{spherevolume}\ee
Using the formula 
\be
\int_0^{\h\pi}d\theta\ (\sin\theta)^{r-1}(\cos\theta)^{s-1}=\h{\Gamma\(\h r\)\Gamma\(\h s\)\over\Gamma\(\h(r+s)\)}\labels{formula}\ee
and \eq{sc},
the integral $\bk{J|J}$ can be calculated for every index set $J$. 

For example, if all the indices in $J$ are equal to $n$, {\it i.e.}, $J=(nnn\cdots n)=(n^p)$, then the integral is
equal to
\be
\bk{J|J}\equiv S(p)&=&{\int_0^{\pi/2}(\cos\theta_1)^{2p}(\sin\theta_1)^{n-2}d\theta_1\over
\int_0^{\pi/2}(\sin\theta_1)^{n-2}d\theta_1}\nn\\
&=&{1\over\sqrt{\pi}}{\Gamma\(p+\h\)\Gamma\(\h n\)\over\Gamma\(p+\h n\)}.\labels{cp}\ee

However, if we replace $J=(n^p)$ by $J=(1^p)$, namely, replacing $x_n^{2p}$
in the integrand by $x_1^{2p}$, then the integral becomes much more harder to calculate,
because $(n-1)$ times more integrations must be performed. 
Yet, on account  of the spherical symmetry, the result must come out to be
the same as \eq{cp}. This complication arises because a choice 
 of axes breaks  the
spherical symmetry. It can be avoided in the invariant
approach discussed below.

This method relies on an explicit parametrization of $\O_{n-1}$ via the spherical coordinates,
as well as formula \eq{formula} allowing the integrations to be carried out. Both
become much more difficult in the $U(n)$ case, so much so that this method is really not very useful
there.
For that reason there shall be no futher discussion of this method.

\subsection{Group theory}
Alternatively, 
$X_J$ can be expanded in terms of spherical harmonics,
and the integral can be transformed into a sum using the orthonormality of
the spherical harmonics. For $n=3$, the expansion is
\be
X_J=\sum_{\ell,m}a_{\ell m}Y_{\ell m}(\theta,\phi),\labels{ylm}\ee
and the integral becomes
\be
\bk{J|J}=\sum_{\ell,m}|a_{\ell m}|^2.\labels{ylm2}\ee
For $n>3$, many more sums are involved in eq.~\eq{ylm2}.

There are two non-trivial tasks in this approach: to find the coefficients
$a_{\ell m}$, and to carry out the  sum in \eq{ylm2}. These tasks become
quite difficult in practice for large $n$ or  $p$.

There is an analogous group-theoretical technique to calculate the $U(n)$ integral,
which is reviewed in Appendix A. Using the Frobenius relation, 
or the Itzykson-Zuber formula, a formula can be
derived to turn the integral into a multiple sum. 
As mentioned in the Introduction, 
 the sums could be very involved,
so in practice this method may not be the best way to obtain a result.
The invariant approach discussed below and in the next section might be simpler.

\subsection{The invariant approach}
The integral in \eq{sphereint} can be calculated directly, using
 only condition \eq{unitsphere} and the invariance of $(d\o)$ under rotation.
In particular,
there is no need to employ the spherical coordinate system, and no need
to know any integration formula, nor group theory. 

The invariant approach will be used in the next section to calculate integrals (of monomials of unitary
matrix elements) over the unitary group $U(n)$. In that case, \eq{unitsphere} is replaced
by the unitarity condition, and $d\o$ is replaced by the invariant Haar measure $dU$ of the unitary group.

It is convenient to arrange the $p$ indices in $J$ according to the 
distinct values (between 1 and $n$) they take.
If $m_1$ of these $p$ indices take on a value, $m_2$ of them take on
a second value,  and so on, then the integral
$\bk{J|J}$ will be denoted by 
 $S(m_1m_2\cdots m_t)$, where $t$ is the number of non-vanishing $m_i$'s, and  $\sum_{i=1}^tm_i=p$. Spherical symmetry guarantees that the integral is independent
of the the specific values the indices assume. This means, among other things,
that  $S$ is symmetrical
in all its arguments.

To calculate $S$, the invariance of $d\o$ 
is used to relate the various $S$'s to $S(p)$.
Then  the value of $S(p)$ is calculated using the sphere condition \eq{unitsphere}.

The invariance of $d\o$ can be exploited in the following way.
A rotation in the $x_ix_k$ plane by an angle $\xi$,
\be
x_i&\to&+c x_i+s x_k\nn\\
x_k&\to&-s x_i+c x_k,\labels{vecrot}\ee
where $c=\cos\xi$ and $s=\sin\xi$,
will leave $d\o$ invariant. Equivalently, if we subject the integrand $X_JX_J$ in $\bk{J|J}$ to such
a rotation, the integral $S(m_1\cdots m_k)$ will remain unchanged.

Let us start out with the integral $S(p)$ whose $X_J$ is equal to $x_1^p$. Under \eq{vecrot}, with $(i,k)=(1,2)$,
the integrand becomes
\be
(x_1^2)^{p}&\to& (cx_1+sx_2)^{2p}=\sum_{e=0}^p{2p\choose 2e}(cx_1)^{2(p-e)}(sx_2)^{2e}+\cdots.\labels{fan1}\ee
The ellipsis indicates terms odd in $x_1$ and $x_2$, which can be dropped because
 they do not contribute to the integral.
The invariance of the integral under this transformation then yields the relation
\be
S(p)=\sum_{e=0}^p{2p\choose 2e}c^{2(p-e)}s^{2e}S(p-e,e).\labels{fan2}\ee
Since this is true for all $\xi$, the right-hand-side must be independent of $\xi$. That requires
\be
S(p-e,e)={{p\choose e}\over{2p\choose 2e}}S(p).\labels{fan3}\ee

Similarly, we can apply \eq{vecrot} and the whole procedure to $(i,k)=(2,3)$ to get
\be
S(p-e,e-f,f)= {{e\choose f}\over{2e\choose 2f}}S(p-e,e)={{p\choose e}\over{2p\choose 2e}}{{e\choose f}\over{2e\choose 2f}}S(p).\labels{fan4}\ee
Continuing thus, we finally obtain
\be
S(m_1,m_2,m_3,\cdots,m_t)={\(\sum_{i=1}^tm_i\)!\over\(\prod_{i=1}^tm_i!\)}{\(\prod_{i=1}^t(2m_i)!\)\over\(\sum_{i=1}^t2m_i\)!}
S(p).
\labels{fan5}\ee

To complete the calculation we must calculate
 $S(p)$. This can be done by using condition \eq{unitsphere}. Since $S(p-1,1)$
is independent of what coordinate $x_j$ the multiplicity 1 sits on, as long as it is not on the coordinate
whose multiplicity is $p-1$, the sphere condition \eq{unitsphere} can be translated to read
\be
(n-1)S(p-1,1)+S(p)=S(p-1).\labels{fan6}\ee
Using \eq{fan5}, we know that $S(p-1,1)=S(p)/(2p-1)$. Substituting this back into \eq{fan6}, we get
the recursion relation
\be
S(p)={2p-1\over n+2p-2}S(p-1).\labels{fan8}\ee
With the initial value $S(0)=1$, \eq{fan8} can be solved to yield
\be
S(p)={(2p-1)(2p-3)\cdots 1\over(n+2p-2)(n+2p-4)\cdots n}={1\over\sqrt{\pi}}{\Gamma\(p+\h\)\Gamma\(\h n\)\over\Gamma\(p+\h n\)},
\labels{fan9}\ee
which agrees with the answer given by \eq{cp}. 

The general result is obtained by substituting \eq{fan9}
into \eq{fan5}.

\section{Integration over the Unitary Group}

Let $U_{ij}$ denote the $(ij)$
matrix element of an $n\x n$ unitary matrix, and $U^*_{ij}$ its complex 
conjugate. 
The product
$\prod_{a=1}^pU_{i_aj_a}$ is abbreviated as $U_{IJ}$, with the index sets being
$I=(i_1i_2\cdots i_p)$ and $J=(j_1j_2\cdots j_p)$.
We shall refer to $p$ as the {\it degree} of $U_{IJ}$.
Since the matrix elements commute with one another, the order of the indices
is irrelevant. Thus if
$P\in S_p$ denotes a permutation of the $p$ indices, and the permuted index set
is denoted by $I_P=(i_{P(1)}i_{P(2)}\cdots i_{P(p)})$, then
\be
U_{IJ}=U_{I_PJ_P}.\label{uu}\ee

We want to calculate the monomial integral
\be
\bk{I,J|K,L}=\int (dU)U^*_{IJ}U_{KL}\labels{int}\ee
over the unitary group $U(n)$. The degree of $U^*_{IJ}$ is 
assumed to be $p$ and 
that of $U_{KL}$
is assumed to be $q$. The Haar measure
$dU$ appearing in \eq{int} is left- and right- invariant, and
normalized to $\int (dU)=1$. 

As mentioned in the last section, a group-theoretical
formula to calculate the integral
is available  (see Appendix A). The integral is expressed as a multiple sum,
with a summand involving the character of the symmetric group $S_p$, the dimensions
of the irreducible representations of $U(n)$ and $S_p$, as well as the index structure
of $I,J,K,L$. This formula is general, though not always the best way to obtain the
result, because the multiple sums are often tedious and difficult to do. In what
follows, we shall develop another method to calculate the integral, using the invariant
approach discussed in the last section. With this approach, no knowledge of group theory
is required, and multiple sums are avoided. 

The invariant calculation relies only on the unitarity of
the matrices in the integrand,  
\be
\sum_{j=1}^nU_{ij}U^*_{lj}=\sum_{j=1}^nU_{ji}U^*_{jl}=\d_{il},\labels{uni}\ee
as well as the invariance of the Haar measure. The latter implies
\be \int (dU)f(U,U^*)=\int (dU)f(VU,V^*U^*)=\int (dU)f(UV,U^*V^*)\labels{inv}\ee 
for any function $f$, and any $V\in U(n)$. We shall apply \eq{inv} to the function
$f(U,U^*)=U^*_{IJ}U_{KL}$.

The calculation is very similar to that
in the last section, though more complicated. The spherical condition \eq{unitsphere} is now
replaced by the unitarity condition \eq{uni}, and the rotational invariance of $d\o$ is now replaced
by the group invariance of $dU$.

\subsection{Relations from invariance}
Eq.~\eq{inv} is very powerful. Depending on the choice of $V$,
many relations can be derived. Here are some examples.

\subsubsection{$V_{ij}=e^{i\phi}\d_{ij}$}
With this choice, $f(UV,U^*V^*)=e^{i(q-p)\phi}f(U,U^*)$. Hence
the integral \eq{int} is zero unless $p=q$. For this reason we shall assume
$q=p$ from now on.

\subsubsection{$V_{ij}=e^{i\phi_i}\d_{ij}$}
With this choice, $f(UV,U^*V^*)=e^{i\xi}f(U,U^*)$, where
$\xi=\sum_{a=1}^p(\phi_{l_a}-\phi_{j_a})$. In order for the integral \eq{int}
not to be zero, this phase $\xi$ must vanish. If all $\phi_i$'s are different,
this happens only when the indices in $L$ are permutations of the indices in $J$.
Similarly, one can show that the indices in $K$ must be permutations of the indices
in $I$. If $R$ and $S$ are the permutations in $S_p$ that do the job, then
\be
K=I_R,\quad L=J_S.\labels{ijkl}\ee
Using \eq{uu}, we may assume $K=I$, with the non-zero integrals \eq{int}
now being of the form 
\be \bk{I,J|I,J_Q}\labels{nonzero}\ee
for some $Q\in S_p$.

The non-zero integral over a sphere calculated in the last section is of the form $\bk{J|J}$, but 
the non-zero integral over $U(n)$ is of the form $\bk{I,J|I,J_Q}$. The presence of an additional
index set $I$, and the possibility that $J_Q\not=J$, both make it harder to calculate the
unitarity integral than the spherical integral, though the idea is precisely the same.

If $J_Q=J$, then the integral \eq{nonzero} is positive definite. We shall refer to integrals
of that type as {\it direct integrals}. If $J_Q\not=J$, the sign is not guaranteed, and we
shall refer to those integrals as {\it exchange integrals}.

The direct integral looks deceptively similar to the spherical integral \eq{sphereint} in
a complex $n^2$-dimensional space. By mapping $U_{ij}$ to $x_a$, with $a=(i,j)$ running between
1 and $n^2$, the complex equivalent of \eq{unitsphere}, namely $\sum_{a=1}^{n^2}|x_a|^2=n$,
is guaranteed by the unitarity relation \eq{uni}. One might therefore think that the direct
integrals would turn out to be very similar to the spherical integrals, whose result is given
by \eq{fan5} and \eq{fan9}. Unfortunately that is not the case, because the measure $dU$
is not rotational invariant in the $n^2$-dimensional complex vector space. As a result, even
the direct integrals become more difficult to calculate than the spherical integrals of the 
last section.

\subsubsection{permutation}
Now, choose $V$ to be a permutation matrix of $n$ objects. Then $VU$ is obtained from
$U$ by permuting its rows, and $UV$ is obtained from $U$ by permuting its columns.
With this $V$, eq.~\eq{inv} implies that
\be
\bk{I,J|K,L}=\bk{I\,',J|K\,',L}=\bk{I,J\,'|K,L\,'},\labels{perm}\ee
where $I\,'$ is obtained from $I$ by a reassignment of the values of its indices, and
$K\,'$ is obtained from $K$ by the {\it same} reassignment. 
For instance, let us take $p=6$ and $n=8$.
Suppose $I=(11555247)$ and  $K=(51524571)$ (recall from \eq{ijkl} that $K$
has to be a permutation of $I$). 
If we make the reassignment $1\leftrightarrow 8,
2 \leftrightarrow 5, 4 \leftrightarrow 7$, then $I\,'=(88222574)$ and $K\,'=(28257248)$.

In other words, the integral is affected by whether the indices 
take on the same or different values, but  is independent of what these values
are.

\subsubsection{row-column interchange}
Since the measure is invariant under transposition,$(dU)=(dU^T)$, the integral
is unaltered if we interchange the rows with the columns:
\be
\bk{I,J|K,L}=\bk{J,I|L,K}.\labels{inter}\ee

\subsubsection{graphical representation}
It is convenient to employ a graphical description of the integral $\bk{I,J|I,J_Q}$,
as illustrated in Fig.~1. The dots in the left-hand column represent
the distinct values in the index set $I$, and the dots in the right-hand column
represent the distinct values of the index set $J$. As per \eq{perm}, the integral
does not depend on what these values are. Since $J_Q$ is a permutation of $J$,
it shares the same distinct values, and hence the same dots as $J$.

 Factors of $U_{IJ}^*$
are shown as (thin) solid lines, and factors of $U_{I,J_Q}$ are shown as dotted lines.
A number appearing above the solid line or below the dotted line denotes the multiplicity
({\it i.e.,} the power) of that matrix element. 
If the number is absent, the multiplicity is taken to be 1.
When a pair $U^*_{ij}U_{ij}$ occurs together, we may 
choose to replace its pair of thin-solid
and dotted lines by a thick solid line. In that case the multiplicity designation
refers to the multiplicity of the pair.
Thus a direct integral can always
be drawn with only thick solid lines.

\vspace*{-1cm}
\begin{figure}[h]
\begin{center}
\includegraphics[bb=87 180 261 381]{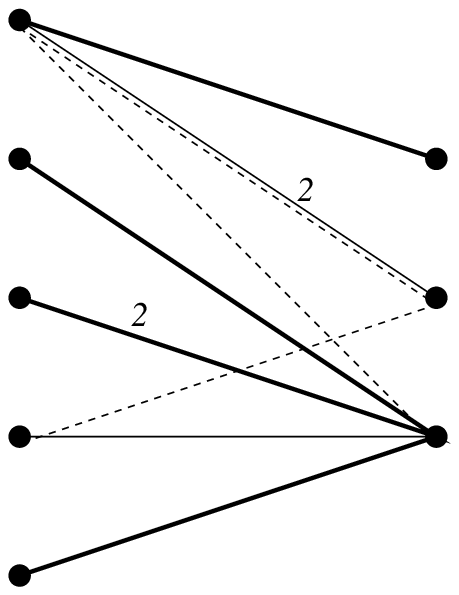}
\caption{Graphical representation of the integral
$\bk{I,J|I,J_Q}$, where $I=(1^323^245), J=(12^27^5)$,
and $J_Q=(217^427)$.}
\end{center}
\end{figure}

When multiplicity is taken into account, 
the number of solid lines and the number of dotted lines connected to each dot
must be equal. This simply reflects the fact that $J_Q$ is a permutation of $J$.

The integral $\bk{I,J|I,J_Q}$ with 
$I=(11123345)=(1^323^245),\ J=(12277777)=(12^27^5)$, and $J_Q=(21777727)=(217^427)$,
is depicted in Fig.~1.
The 5 left-hand dots represent the five distinct numbers 1,2,3,4,5
in $I$, and the 
three dots in the right-hand column represent the three distinct numbers 1,2,7 in $J$
and $J_Q$. 

The invariance of \eq{inter} means that we may switch the left-hand column of
 dots with the right-hand column of dots.
Namely, a reflection about the
vertical line halfway between the two columns will not change the integral.

\subsubsection{rotation}
Choose $V=R(ab)$ to be the matrix which rotates the  $(ab)$ plane by an angle
$\xi$. This matrix has 1's along the main diagonal, except at the $(aa)$ and
$(bb)$ positions, where the matrix element is $c=\cos\xi$. The off-diagonal
matrix elements are all zero, except at the positions $(ab)$ and $(ba)$, where the
matrix elements are respectively $s=\sin\xi$ and $-s$.

 The replacement
$U\to UV$ causes the following change in the matrix elements:
\be
U_{ia}&\to& +c\ U_{ia}+s\ U_{ib},\nn\\
U_{ib}&\to& -s\ U_{ia}+c\ U_{ib},\nn\\
U_{ij}&\to &U_{ij},\labels{uz1}\ee
provided $j\not= a,b$. Similar replacements on $U^*_{ia}$ and $U^*_{ib}$
should also be made. The result is to change $\bk{I,J|I,J_Q}$ into 
a sum of terms of the form\footnote{Odd powers of $cs$ never enters because of \eq{ijkl}.}
 $M_e(c^2)^{d-e}(s^2)^e$, where $d$ is the total number
of column indices in $U^*_{IJ}$ with value $a$ or $b$, and $e$ varies
between 0 and $d$. The invariance condition \eq{inv} requires
\be
\bk{I,J|I,J_Q}=\sum_{e=0}^dM_e(c^2)^{d-e}(s^2)^e.\labels{rot}\ee
In order for this to be true for all $\xi$, we must have
\be
M_e=\bk{I,J|I,J_Q}{d\choose e}=M_0{d\choose e},\labels{binom}\ee
where ${d\choose e}=d!/e!(d-e)!$ is the binomial coefficient.

Let us see how $M_e$ is computed in the graphical language. Take any two dots on
the right-hand column to represent the values $a$ and $b$. 
One of the two dots should have some lines attached to it, but
the other one may or may not be empty. 
The total number of solid (or of dotted) lines attached to the two dots is $d$.
Now move $e$ (thin) solid and $e$ dotted lines 
between the two dots, subject to the constraint that at the end of the move, each
dot must have an equal number of solid and dotted lines attached to it
(otherwise the integral is zero). Assign a weight $+1$ for a line moved from $a$ to 
$b$, and a weight $-1$ for each line moved from $b$ to $a$. 
The quantity $M_e$ in \eq{binom} is simply the sum of
all the resulting integrals after the move, weighted by the product of the factors $\pm 1$ associated with
each move. 

It is important to note that these relations are local. They involve only the indices
of $J$ and $J_Q$ with values $a$ and $b$. It does not matter what $I$ is and what the rest of 
the indices of $J$ and $J_Q$ are.
 
Let us illustrate this graphical application of the rotational relation with two examples.

\bigskip
\n\underline{\bf example 1}

Fig.~2 represents part of a diagram. The whole diagram may have
 many more dots and lines. They are
not drawn because the relation derived below is independent of these other dots and lines.

\begin{figure}[h]
\begin{center}
\includegraphics[bb=20 0 520 400,scale=0.8]{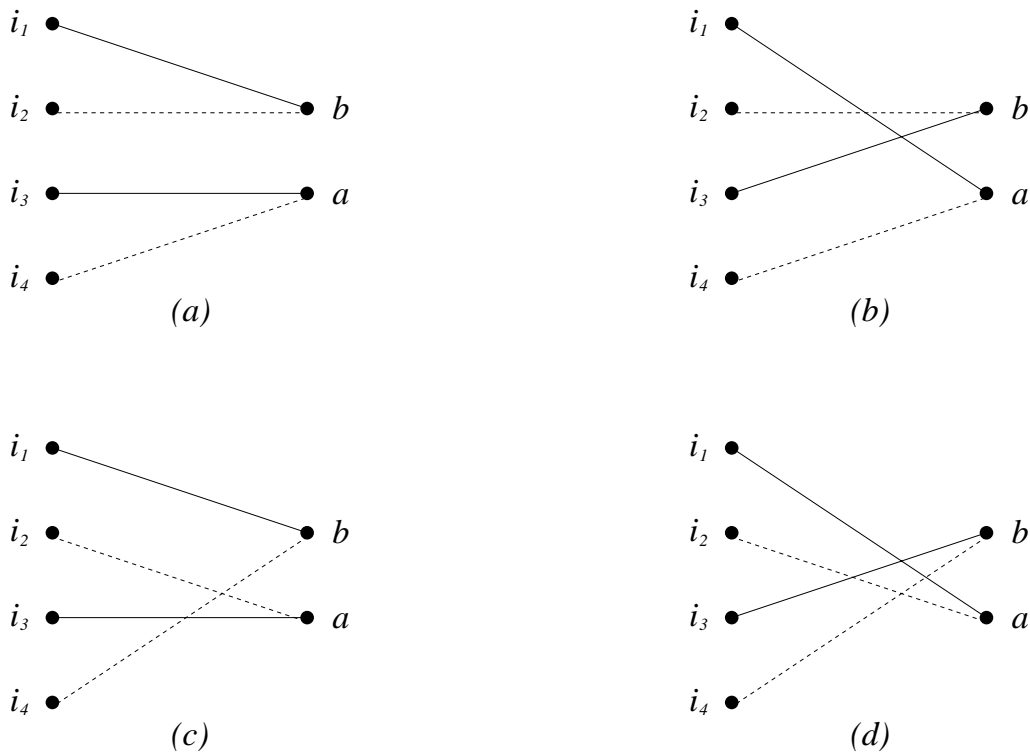}
\caption{Four diagrams illustrating how rotational invariance of the
measure can be used to relate these four integrals. See eq.~\eq{rot2}.}
\end{center}
\end{figure}

The values of $a$ and $b$ must be different, and there
must be no more lines coming out of them
than is shown in Fig.~2.
The indices $i_1,i_2,i_3,i_4$ may or may
not have the same values, and there may be many more lines attached to them than is 
shown, as long as these other lines do not connect with either $a$ or $b$.  
If some of their values are the same, say $i_1=i_2$, then graphically
the two dots $i_1,i_2$ simply merge together into a single dot. If the values
are different, then other lines must also come out of these dots in order to make the integral
non-zero. 

The integrals corresponding to Figs.~2(a), 2(b), 2(c), and 2(d)
will be denoted by $I(2a)$, $I(2b)$, $I(2c)$, and $I(2d)$,
respectively. In these diagrams 
we are dealing with $d=2$, because there are two pairs of (solid and dotted) 
lines ending on
the right-hand dots.

Let us apply rotation to Fig.~2(a). We could either move two lines ($e=1$),
or four lines ($e=2$). By moving two lines, we get $I(2b)$ (two solid lines), $I(2c)$ (two dotted lines),
and two others, $I_1$ (one solid and one dotted lines from $a$ to $b$) and $I_2$ (one solid and one
dotted lines from $b$ to $a$). The graphs for the last two are not shown, but they can be obtained 
from $I(2a)$ by merging the $a$ and $b$ dots. 
In this way we get $M_1=
-I(2b)-I(2c)+I_1+I_2=2(-I(2b)+I_1)$, where \eq{perm} has been used in the last step.
 By using \eq{binom}, we conclude that $M_1=2M_0=2I(2a)$. Hence we obtain the relation
\be
I(2a)=-I(2b)+I_1.\labels{rot2}\ee

If we move all four lines, we get $M_2=I(2d)$. The formula in \eq{binom}
demands $M_2=M_0$, or $I(2d)=I(2a)$. We already know this to be true from \eq{perm}.

\bigskip
\n\underline{\bf example 2: the fan relation}

Fig.~3(b) is a partial diagram of some integral. We assume there are no other
lines connected to the 
dots on the right-hand column, though there may be other lines emerging
from dot $i$. There may also be many other dots and lines not shown in the diagram.

\vspace*{-1cm}
\begin{figure}[h]
\begin{center}
\includegraphics[bb=40 40 520 330]{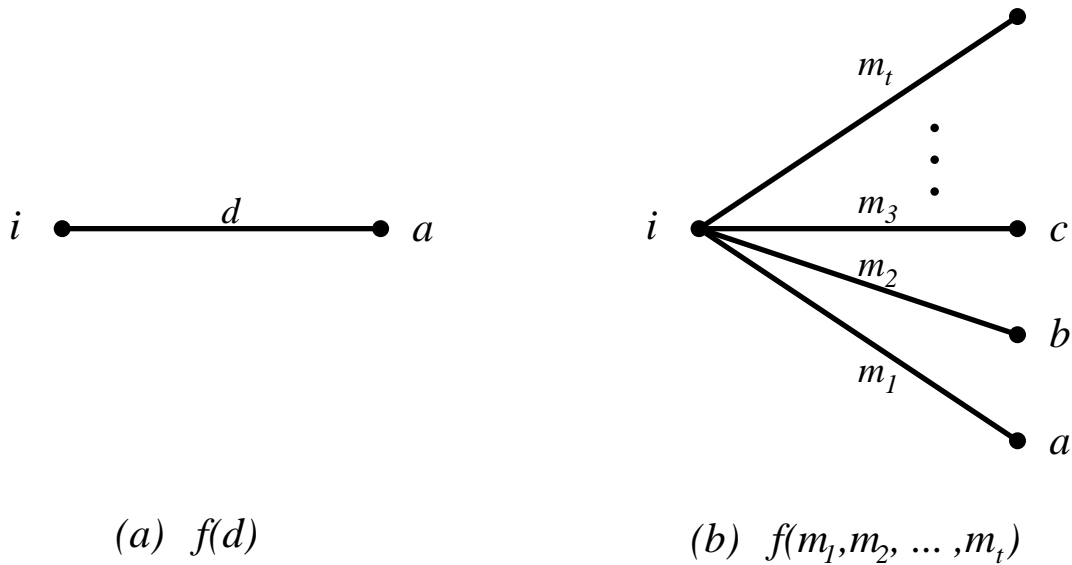}
\caption{Using rotational invariance of the measure, the single
line in $(a)$ can be spread out into the fan of lines in $(b)$
to get the relation displayed in \eq{rot5}.}
\end{center}
\end{figure}

The multiplicities of the lines shown are $m_i\ (1\le i\le t)$, 
so the integrand is proportional 
to $|U_{ia}|^{2m_1}|U_{ib}|^{2m_2}\cdots$.
The corresponding integral is denoted by $f(m_1,m_2,\cdots,m_t)$.

Using \eq{binom} repeatedly, it will be   shown below that
\be
f(m_1,m_2,\cdots,m_t)={\(\prod_{i=1}^tm_i!\)\over \(\sum_{i=1}^tm_i\)!}f(d),\labels{rot5}\ee
where $d=\sum_{i=1}^tm_i$ and $f(d)=f(d,0,0,\cdots,0)$. 

$f(d)$ is drawn in Fig.~3(a). Having multiplicity $d$ means that there are $d$ (thin) solid lines
and $d$ dotted lines between $i$ and $a$. Now move 
$e=\sum_{i=2}^tm_i$ solid and $e$ dotted lines from $a$ to an empty
dot $b$. There are ${d\choose e}$ ways of choosing the set of solid lines to move, and independently
there are also ${d\choose e}$ ways to select the dotted lines. Hence $M_e={d\choose e}^2f(d-e,e)$.
From \eq{binom}, we know that $M_e={d\choose e}M_0={d\choose e}f(d)$. Hence
\be
f(d-e,e)={1\over{d\choose e}}f(d).\labels{b1}\ee

This process can be repeated by moving $g=\sum_{i=3}^tm_i$ pair of lines from $b$ to an empty dot $c$.
Then we get
\be
f(d-e,e-g,g)={1\over{e\choose g}}f(d-e,e)={1\over{d\choose e}}{1\over{e\choose g}}f(d).\labels{b2}\ee
By repeating this process again and again, we  arrive at the fan relation \eq{rot5},
which tells us how to fan out a thick line with a high multiplicity
  into $t$ different lines.

\subsection{Results from unitarity}
In the last subsection, relations between different integrals are obtained
using the invariance requirement \eq{inv}.
To calculate the actual value of any of these integrals, the unitarity condition
\eq{uni} must be used.

The unitarity sum \eq{uni} for $i\not=l$ simply brings out more relations between different integrals.
But for $i=l$, a pair of $U,U^*$ disappears on the right-hand side of $\eq{uni}$, so 
\eq{uni} relates integrals of degree $p$ to integrals of degree $p-1$. By using this repeatedly, eventually
the degree comes down to  zero, and the integral is known to be 1. In this way 
 the values of the integrals
can be computed recursively.

This procedure will be illustrated by various examples in 
the rest of this section.

\subsubsection{direct integrals}

\n\underline{\bf the fan integrals}

The simplest ($p=1$) direct integral is 
\be
\bk{i,a|i,a}={1\over n}\labels{1}\ee
To get this result, we make use of the fact from
\eq{perm} that $\bk{i,a|i,a}$ is independent of $i$ and $a$.
Summing over $a$ (from 1 to $n$), and using \eq{uni}, we get $n\bk{i,a|i,a}=1$.
Hence eq.~\eq{1}.

This calculation can be generalized to the integral in Fig.~4(a) to give
\be
\bk{(i^m),(j^m)|(i^m),(j^m)}\equiv F(m)={(n-1)!m!\over(n+m-1)!}.
\labels{4a}\ee
\break
Note that Fig.~4(a) is just the diagram  Fig.~3(a), but without any additional
dots and lines.

\vspace*{-1cm}
\begin{figure}[h]
\begin{center}
\includegraphics[bb=40 480 600 800]{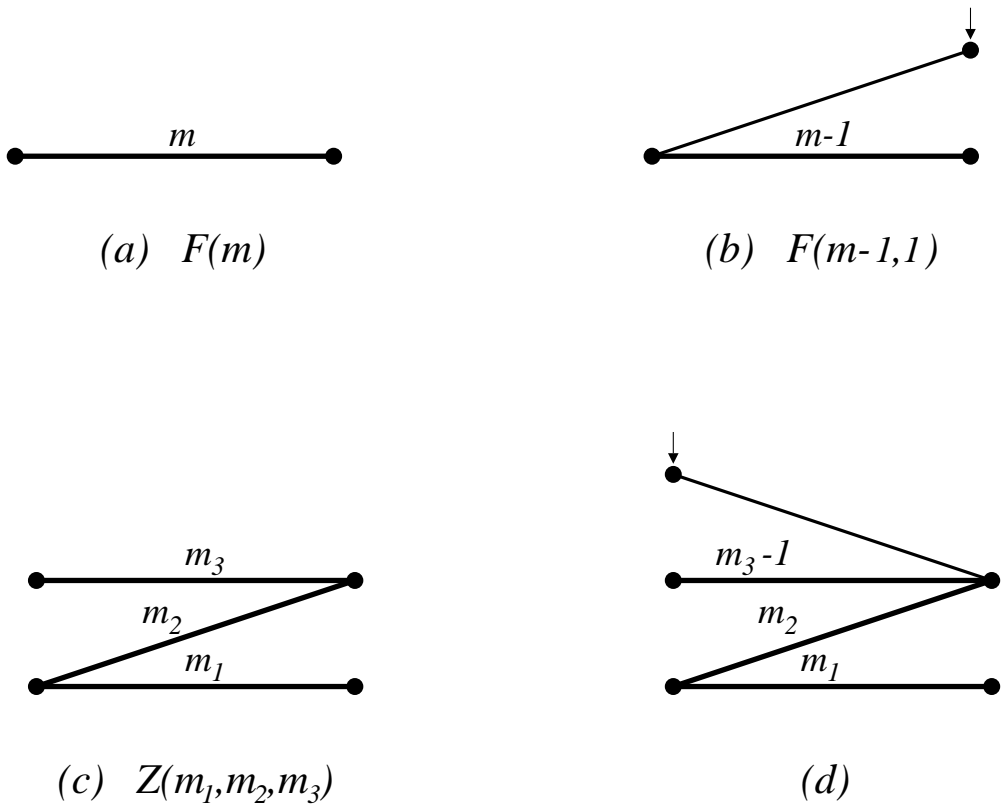}
\vspace*{1cm}
\caption{Diagram $(b)$ shows  how the integral in $(a)$ can be obtained from unitarity
by summing over the index indicated by an arrow. A recursion relation
whose solution is  \eq{4a}.
Similarly, unitarity applied to the indicated index in (d) yields a recursion
relation whose solution gives \eq{4c} for the `Z integral'  
in (c).}
\end{center}
\end{figure}

To obtain this result, start from the integral
$F(m-1,1)$, shown in Fig.~4(b). The integral is independent of the value
of the index (dot)  
indicated by a downward arrow, as long as it does not take on
the value of the other dot.  Sum over this indicated index, from 1 to $n$.
Eq.~\eq{uni} implies 
\be
(n-1)F(m-1,1)+F(m)=F(m-1).\labels{4b}\ee
Now the fan relation \eq{rot5} tells us that $F(m-1,1)=F(m)/m$.
Substituting this into \eq{4b}, we get a recursion relation between $F(m)$ and $F(m-1)$,
namely, 
\be
F(m)=F(m-1){m\over n+m-1}.\labels{4a3}\ee
Using the initial value $F(0)=1$, this recursion relation can be solved to get \eq{4a}.

Define the `fan integral' 
$F(m_1,m_2,\cdots,m_n)$ to be Fig.~3(b), without any extra dots and lines.
It follows from \eq{4a} and \eq{rot5} that
\be
F(m_1,m_2,\cdots,m_t)={\(\prod_{i=1}^tm_i!\)(n-1)!\over\(n+\sum_{i=1}^tm_i-1\)!}.\labels{4a2}\ee

\n\underline{\bf the $Z$ integrals}

Next, consider the `Z integral' in Fig.~4(c). We shall prove that
\be
Z(m_1,m_2,m_3)={m_1!m_2!m_3!(n-2)!(n-1)!(n+m_1+m_3-2)!\over
(n+m_1-2)!(n+m_3-2)!(n+m_1+m_2+m_3-1)!}.\labels{4c}\ee
To do so, consider Fig.~4(d).
Summing over the index indicated by the vertical arrow, and denoting the integral
in Fig.~4(d) by $I(4d)$, the unitarity condition \eq{uni} gives
\be
(n-2)I(4d)+Z(m_1,m_2,m_3)+Z(m_1,m_2+1,m_3-1)=Z(m_1,m_2,m_3-1).\labels{4c2}\ee
The fan formula \eq{rot5} tells us that $I(4d)=Z(m_1,m_2,m_3)/m_3$. Substituting this
into \eq{4c2}, we get a recursion relation in $m_3$:
\be
Z(m_1,m_2,m_3)={m_3\over n+m_3-2}\[Z(m_1,m_2,m_3-1)-Z(m_1,m_2+1,m_3-1)\].\labels{4c3}\ee
Using the initial value $Z(m_1,m_2,0)=F(m_1,m_2)=(n-1)!m_1!m_2!/(n+m_1+m_2-1)!$,
the recursion relation can be solved to yield \eq{4c}.  

Relation \eq{rot5} can also be used to fan out the two open ends of Fig.~4(c) and \eq{4c} 
to obtain the `fanned Z integrals'.

\subsection{Exchange integrals}
To illustrate how to compute exchange integrals, 
all second and third degree exchange integrals will be
computed in this subsection.

\subsubsection{$p=2$}
All the second degree integrals are shown in Fig.~5. The integrals in Figs.~5(a), 5(c), 5(d), 5(e) are direct
integrals, either of the fan type, or the Z type, so they are known. The only exchange integral is $E(2)$ depicted
in Fig.~5(b). It can be computed either by rotation from a direct integral, or by unitarity. We will discuss
both methods.

\vspace*{-3cm}
\begin{figure}[h]
\begin{center}
\includegraphics[bb=0 0 540 500,scale=0.9]{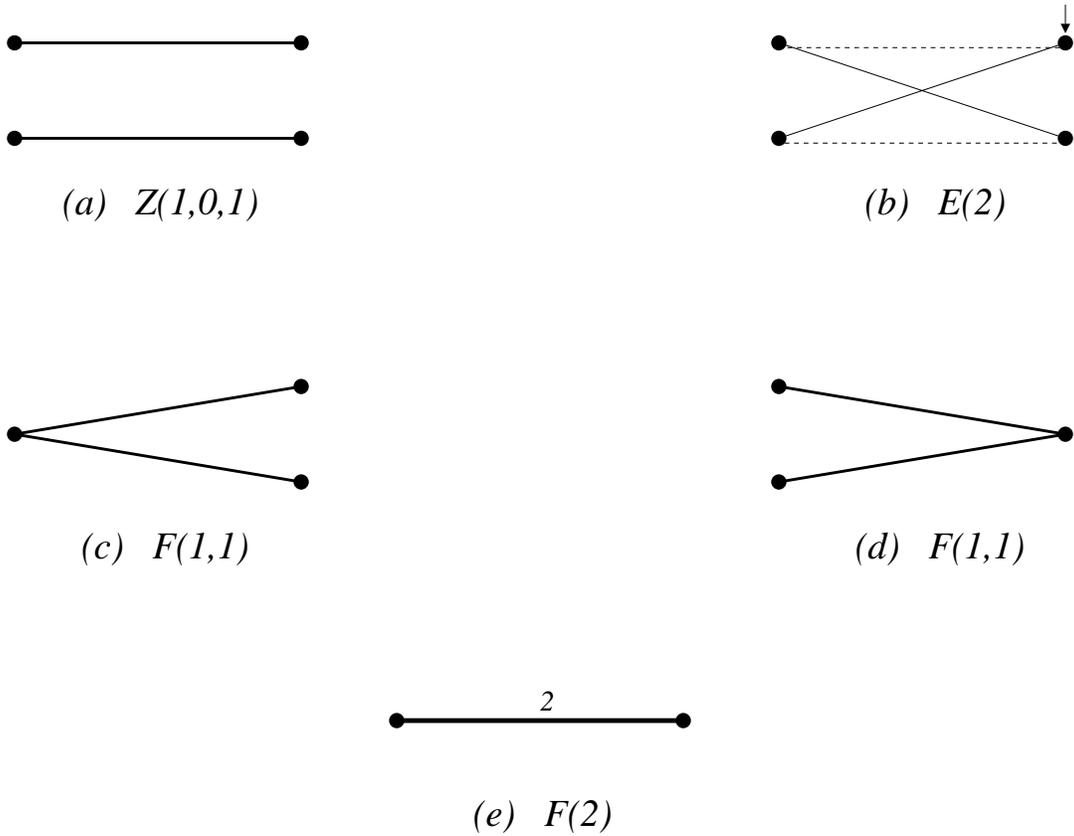}
\caption{Integrals of the second degree.}
\end{center}
\end{figure}

\n\underline{\bf by rotation}

Starting from Fig.~5(d), rotate two solid and two dotted lines from the 
dot in the right-hand column
 to an empty dot. This is a special
case of example 1 of Sec.~IIIA5 and Fig.~2, but let us do it directly
once again. Using the notation in \eq{binom}, we get
\be
M_1=2Z(1,0,1)+2E(2)=2M_0=2F(1,1).\labels{e21}\ee
Hence
\be
E(2)=F(1,1)-Z(1,0,1)={1\over n(n+1)}-{1\over (n-1)(n+1)}=-{1\over n(n^2-1)}.\labels{e22}\ee

\n\underline{\bf by unitarity}

Summing over the indicated index  in Fig.~5(b) from 1 to $n$ yields $(n-1)E(2)+F(1,1)=0$, hence
\be
E(2)=-{F(1,1)\over n-1}=-{1\over n(n^2-1)}.\labels{e23}\ee

\subsubsection{$p=3$}
Fig.~6 shows the two direct integrals which are not of the fan type or the Z type, and all the exchange integrals
of degree 3.

To get $I(6a)$, the integral for Fig.~6(a), carry out a unitarity sum on the indicated index. This yields
$(n-1)I(6a)+F(1,1,1)=F(1,1)$, hence
\be
I(6a)={1\over n-1}\({1\over n(n+1)}-{1\over n(n+1)(n+2)}\)={1\over (n-1)n(n+2)}.\labels{6a}\ee
This integral
can also be computed by fanning out the bottom line of $Z(2,0,1)=2(n+ 1)(n- 2)!/(n+2)!$.

To compute $I(6b)$, take a unitary sum over the indicated index. This yields $(n-2)I(6b)+2I(6a)=Z(1,0,1)$. Hence
\be
I(6b)={1\over n-2}\({1\over n^2-1}-{2\over (n-1)n(n+2)}\)={(n^2-2)(n-3)!\over(n+2)!}.\ee

\vspace*{1cm}
\begin{figure}[h]
\begin{center}
\includegraphics[bb=0 300 540 800, scale=0.8]{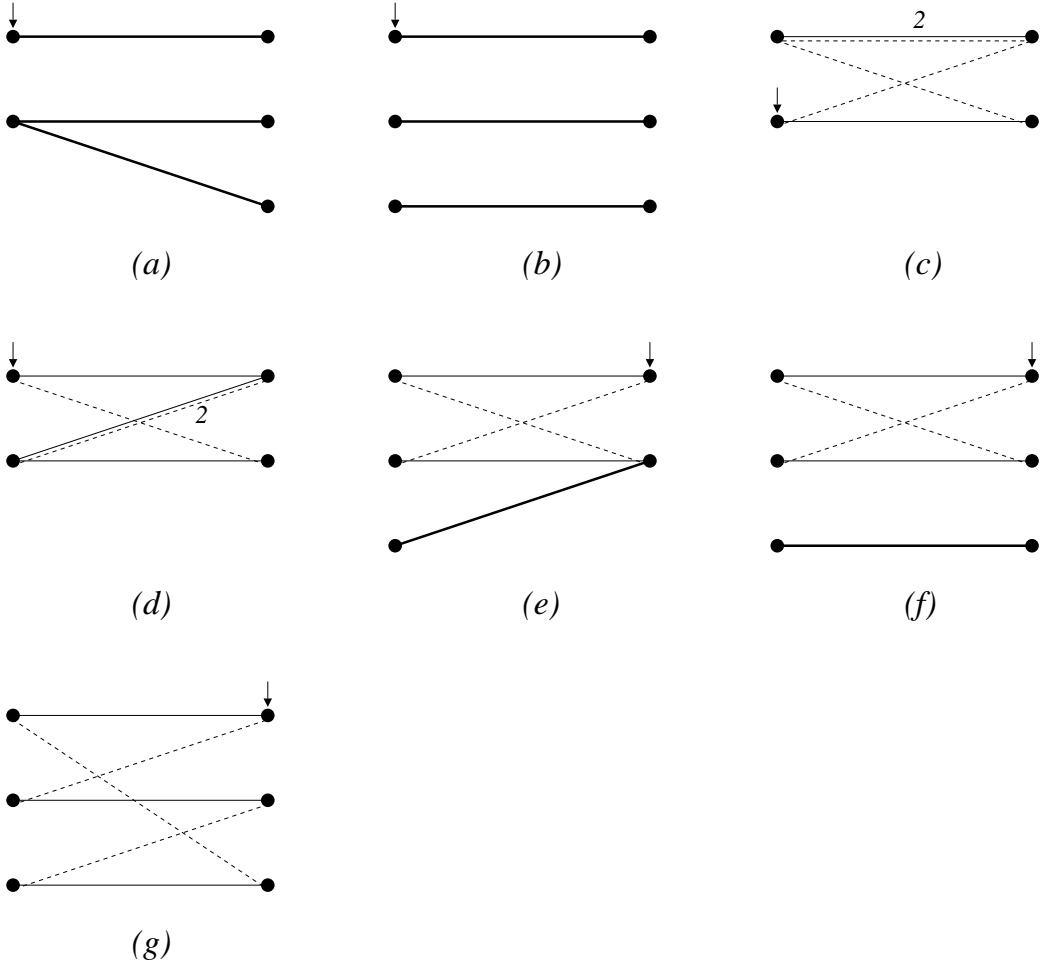}
\caption{Integrals of the third degree.}
\end{center}
\end{figure}

The exchange integrals can be obtained by taking the unitary sum on the indicated
vertex. Note that the right-hand side of the sum is always zero 
in the case of exchange integrals.

In this way we obtain the relations
\be
(n-1)I(6c)+F(1,2)&=&0\nn\\
(n-1)I(6d)+F(1,2)&=&0\nn\\
(n-1)I(6e)+F(1,1,1)&=&0\nn\\
(n-2)I(6f)+I(6a)+I(6e)&=&0\nn\\
(n-2)I(6g)+2I(6e)&=&0.\labels{61}\ee
The solutions are
\be
I(6c)=I(6d)&=&-{2(n-1)!\over(n+2)!(n-1)}=-2{(n-2)!\over (n+2)!}\nn\\
I(6e)&=&-{F(1,1,1)\over n-1}=-{(n-2)!\over(n+2)!}\nn\\
I(6f)&=&-{1\over n-2}\(I(6a)+I(6e)\)=-n{(n-3)!\over (n+2)!}\nn\\
I(6g)&=&-{2\over n-2}I(6e)=2{(n-3)!\over (n+2)!}.\labels{62}\ee

\subsection{The X integrals}
\vspace*{-2cm}
\begin{figure}[h]
\begin{center}
\includegraphics[bb=20 0 530 600,scale=0.6]{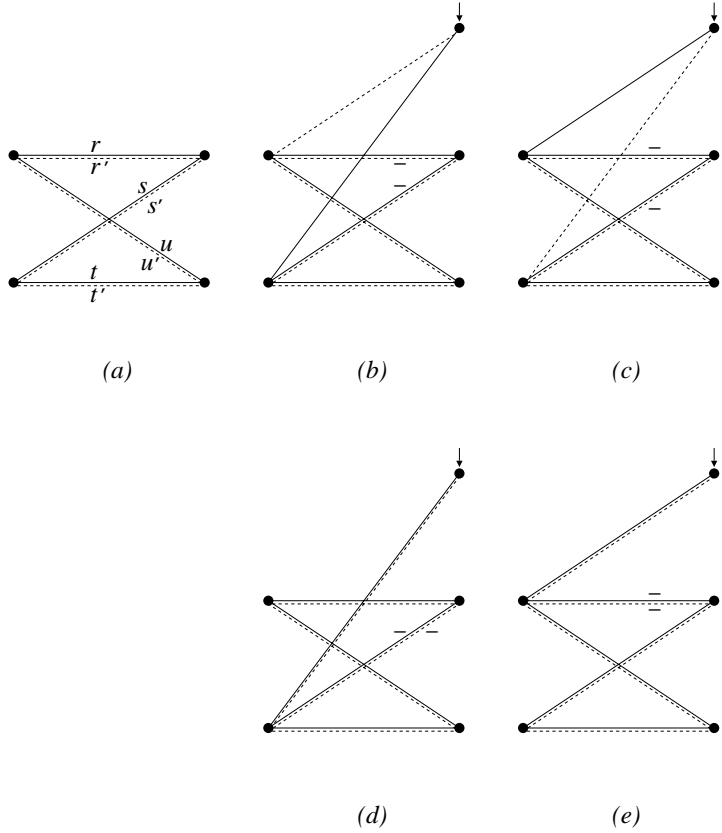}
\caption{($a$) The `X integral' and its weights. $(b)-(e)$ Diagrams 
obtained from ($a$) by rotating away a pair of lines from the $r-s$ junction.
The weights in these four diagrams are the same as those in $(a)$,
except for the ones with a `$-$' sign, in which case the corresponding
weight is decreased by 1.}
\end{center}
\end{figure}

To illustrate how direct and exchange integrals may be coupled
in the recursion relation, let us look at
the 1-loop `X integrals' depicted in Fig.~7(a). This integral, specified by
the four weights $r,s,t,u$ of $U^*$ and the four weights $r',s',t',u'$ of $U$,
will be designated as $X(rstu|r's't'u')=I(7a)$.
Since the number of dotted lines and the number of solid lines emerging from
each vertex must be equal, there are three relations for these
eight parameters,
\be
r'+s'&=&r+s,\nn\\
s'+t'&=&s+t,\nn\\
t'+u'&=&t+u,\labels{x0}\ee
so only five independent parameters are required to specify all the $X$ integrals.

It is fairly complicated to calculate all these integrals, so we will only
derive the recursion relation here and illustrate how it can be used in the simplest case.
Let $\rho=r+r'$ be the total number of top lines in $X$,
and $\sigma=s+s'$ the total number of lines on one side.
The idea is to find a recursion relation in $\rho+\sigma$, each time reducing
either $\rho$ or $\sigma$ by 1.
Eventually one gets down to either $\rho=0$ or $\s=0$, 
which are the $Z$ integrals obtained before. 

\subsubsection{recursion relation}
Rotate one solid and one dotted line from the $r-s$ junction of Fig.~7(a)
to an empty dot. The result is Fig.~7(b) to Fig.~7(e). In the notation of
\eq{binom}, we have 
\be
M_1=r'sI(7b)+rs'I(7c)+ss'I(7d)+rr'I(7e)=(r+s)M_0=(r+s)I(7a),\labels{x1}\ee
where $I(7b)$ is the integral depicted in Fig.~7(b), etc.  

The parameters of Figs.~7(b) to 7(e) are those of 7(a), except where a
`$-$' sign occurs, in which case the corresponding parameter is reduced by 1.
In Fig.~8 we will also use a `$+$' sign to indicate where
the parameter is increased by 1.

The unitarity sum, applied 
to the indicated index in Figs.~7(b) to 7(e), results in the relations
\be
0&=&(n-2)I(7b)+I(7a)+I(8b)\nn\\
0&=&(n-2)I(7c)+I(7a)+I(8c),\nn\\
I(8d_2)&=&(n-2)I(7d)+I(7a)+I(8d_1),\nn\\
I(8e_2)&=&(n-2)I(7e)+I(7a)+I(8e_1).\labels{x2}\ee
A substitution of \eq{x2} into \eq{x1} yields the desired recursion relation
\be
I(7a)=-{r'sI(8b)+rs'I(8c)+ss'[I(8d_1)-I(8d_2)]+rr'[I(8e_1)-I(8e_2)]
\over(r+s)(r'+s'+n-2)}.\labels{x3}\ee

\begin{figure}[h]
\begin{center}
\includegraphics[bb=0 0 540 500,scale=0.6]{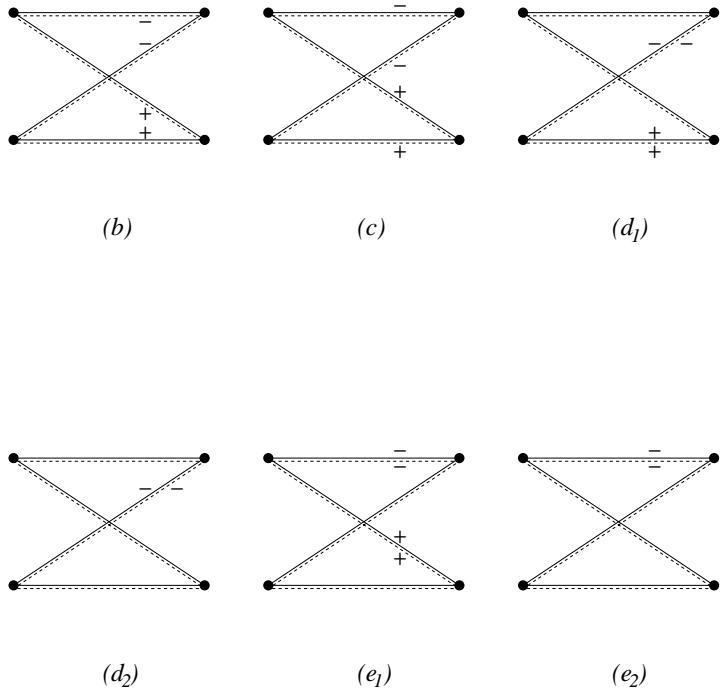}
\caption{Terms obtained from the unitarity sum of Figs.~$7(b)$ to $7(e)$.
The weights in these diagrams are the same as in Fig.~7(a), except on
those lines indicated by a `$-$' or a `$+$'. In those cases, the
corresponding weights are decreased or increased by 1.}
\end{center}
\end{figure}

Let us illustrate the recursion relation by computing the simplest cases, with 
$\rho+\s=2$.

There are four possibilities with $r+r'+s+s'=2$. Two of them are the
known $Z$-integrals, namely, $X(0,1,t,u|0,1,t,u)=Z(1,t,u)$ and $X(1,0,t,u|
1,0,t,u)=Z(1,u,t)$. The other two are exchange integrals which can
be obtained from \eq{x3}:
\be
X(1,0,t,u|0,1,t-1,u+1)&=&-{I(8c)\over n-1}=-{F(t,u+1)\over n-1}=-{t!(u+1)!(n-2)!\over (n+t+u)!},\labels{x4}\\
X(0,1,t-1,u|1,0,t,u-1)&=&-{I(8b)\over n-1}=-{F(t+1,u)\over n-1}=-{(t+1)!u!(n-2)!\over (n+t+u)!}.\labels{x5}\ee
In particular, \eq{x4} reduces to $I(6d)$ of \eq{62} when $t=u=1$.

\section{Conclusion}
We have shown how an integral over the $U(n)$ manifold can be computed recursively 
using only
the invariance of the Haar measure and the unitarity condition. The same method can
also be used to get a closed expression for a monomial integral over a unit sphere.

In a forthcoming paper, we shall compare the relative advantages of the invariant method
developed here, and the group-theoretical method reviewed in Appendix A. We will
show how the two methods can be combined to strengthen each other.

CSL would like to thank G. Semenoff and A. D'Adda for helpful discussions.
This research is supported by the Natural Sciences and Engineering Research
Council of Canada and by the Fonds de recherche sur la nature et les
technologies of Qu\'ebec.

\appendix
\section{Group Theoretical Calculation}
As shown in \eq{nonzero}, the non-zero integrals of \eq{int} can be
written in  the form $\bk{I,J|I,J_Q}$. Using group theory, to be reviewed
below, the integral can be turned into a 
multiple sum\footnote{CSL wishes to thank Prof.~Alessandro D'Adda for introducing him to this formula.}:
\be
\bk{I,J|I,J_Q}=\sum_{R\in{\cal G}_I}\sum_{S\in{\cal G}_J}\sum_f{d_f^2\over
(p!)^2\b d_f}\chi_f(SQR),\labels{gpint}\ee
where $p$ is the degree of $U_{IJ}$
appearing in the integral \eq{int}. The other symbols will be explained below.

The irreducible representations of the unitary group $U(n)$ 
are labeled by a Young's tableau. It is defined by
a sequence of non-negative integers $f=(f_1f_2\cdots f_n)$, with
$f_i\ge f_{i+1}$. All irreducible representations contained in
a $p$th rank tensor have their Young's tableaux restricted to 
$p$ boxes, namely, $\sum_{i=1}^nf_i=p$. In that case 
clearly $f_i=0$ for $i>p$. It is 
customary to drop the zeros when the sequence $f$ is written,
so $f$ can be written as $(f_1f_2\cdots f_p)$, or even shorter if there
are more zeros.

The dimension of the irreducible representation $f$ of $U(n)$, denoted by $\b d_f$,
is given by the ratio of two Vandermonde determinants
\be
\b d_f={D(\ell_1,\ell_2,\cdots,\ell_n)\over D(n-1,n-2,\cdots,0)},\labels{dimun}\ee
where $\ell_i=f_i+n-i$, and
\be
D(x_1,x_2,\cdots,x_n)=\prod_{i>j=1}^n(x_i-x_j).\labels{vdm}\ee

The irreducible representations of the symmetric group $S_p$ are also labeled
by  Young's tableaux $f=(f_1f_2\cdots f_p)$.
The dimension of an irreducible representation of $S_p$ is denoted by $d_f$, and 
the character for the element $P \in S_p$ is denoted by $\chi_f(P)$.
Tables are available to give their values for small $p$.

The character $\chi_f(P)$ depends only on the class that $P$ belongs to. If 
a permutation $P$
is written in the cycle form, then permutations with the same cycle
structure belong to the same class. The cycle structure can be labeled by
a Young's tableau $c=(c_1c_2\cdots c_p)$, where $c_1$ is the length of the longest
cycle in $P$, $c_2$ is the length of the next longest cycle in $P$, etc. If
$P\in c$, we will also write $\chi_f(P)$ as $\chi_f(c)$. The characters together
are given by $\gamma_p^2$ numbers, where $\gamma_p$ is either the total number of distinct
classes in $S_p$, or the number of inequivalent irreducible representations. 
It is equal to the number of partitions of $p$, and is given by 
$\gamma_p=1,2,3,5,7,11,15$, for $p=1,2,3,4,5,6,7$, respectively.

${\cal G}_I\subset S_p$ is the symmetry group of the index set $I$, and
${\cal G}_J\subset S_p$ is the symmetry group of the index set $J$.
For example, if $I=(111338888)$, then ${\cal G}_I=S_3\x S_2\x S_4\subset S_9$.
If $I=(13254798)$, then ${\cal G}_I$ consists of the identity $e$ only.

The sum in \eq{gpint} is over all the irreducible representations $f$, all elements $R$ of the
symmetry group ${\cal G}_I$, and all elements $S$ of the symmetry group ${\cal G}_J$.

The simplest integrals to calculate are those where the indices in $I$ all take on distinct values, and 
similarly for $J$. In that case, ${\cal G}_I={\cal G}_J=e$, so the sums in \eq{gpint} reduces
the single sum over the irreducible representations $f$. Since $\chi_f(Q)$ depends on the 
class $c$ $Q$ lies in, there are $\gamma_p$ distinct integrals $\bk{I,J|I,J_Q}$ of this type.
It is convenient to denote these integrals by $\xi(c)$.
 To compute them, we need to use \eq{dimun} to compute $\b d_f$, a character
table of $S_p$ to compute $\chi_f$ and $d_f$, then we must sum up $\gamma_p$ terms
in \eq{gpint} before we get $\xi(c)$.

More generally,
\eq{gpint} can be written in terms of $\xi(c)$ as
\be
\bk{I,J|I,J_Q}=\sum_c N(I,J,Q|c)\xi(c),\nn\labels{gp2}\ee
where
\be
N(I,J,Q|c)=\sum_{R\in{\cal G}_I}\sum_{S\in{\cal G}_J}\d(SQR\in c)\labels{gp4}\ee
is the total number of $SQR$ 
in class $c$. They are often quite tedious to compute. Once it is
calculated, we still have to carry out the $\gamma_p$ sums over $c$ to get $\bk{I,J|I,J_Q}$.

This completes the description of formula \eq{gpint}. 
In the remainder of this Appendix, we will sketch 
how it is arrived at.

The orthonormal relation for the irreducible representaions $D_f(R)$ of the
$S_p$ group is
\be
{1\over f!}\sum_{P\in S_p}\[D_f(P)\]_{il}\[D_g(P^{-1})\]_{mj}=
{1\over d_f}\d_{fg}\d_{ij}\d_{lm}.\labels{f2}\ee
This leads to the following relation for characters, true for any $Q$ and $R$ in $S_p$;
\be
{1\over p!}\sum_{P\in S_p}\chi_f(PQ)\chi_g(RP^{-1})&=&\d_{fg}{1\over d_f}\chi_f(QR).\labels{f3}\ee
The corresponding character relation for $U(n)$,
\be
\int (dU)\b\chi_f(UV)\b\chi_g(WU^{-1})&=&\d_{fg}{1\over \b d_f}\b\chi_f(VW),\labels{f4}\ee
is true for any $V$ and $W$ in $U(n)$.

Given a $U\in U(n)$ and a  $P\in S_p$, define
$(U)_P$ to be $\sum_IU_{I\,I_P}$, where the sum is taken over all the indices
in the set $I=(i_1i_2\cdots i_p)$, each covering its full range 
of values from 1 to $n$. The index set
$I_P$ as well as $U_{I\,I_P}$ are defined at the beginning of Sec.~III. If $P$ consists
of $\alpha_i$ cycles of length $i$, then
\be
(U)_P=\prod_j\(\Tr(U^i)\)^{\alpha_i}.\labels{up}\ee
Since $(U)_P$ depends only on the cycle structure of $P$, it is the same for two $P$'s
in the same class.

The crucial input to the computation of the integral is Frobenius formula,
\be
(U)_P=\sum_f\b\chi_f(U)\chi_f(P).\labels{frobenius}\ee
Applying it to $(UV)_e$ and $(WU^{-1})_e$, and using \eq{f4} to
integrate, one arrives at the expression
\be
\int (dU)(UV)_e(WU^\dagger)_e=\sum_f{d_f^2\over
\b d_f}\b\chi_f(VW),\labels{f6}\ee
where $d_f=\chi_f(e)$ has been used.
Next, use \eq{f3} to introduce the factor
\be
\d_{fg}={1\over p!}\sum_{P\in S_p}\chi_f(P)\chi_g(P^{-1})\labels{f7}\ee
to the right-hand side of \eq{f6}, and use the Frobenius formula again (note 
that $\chi_g(P^{-1})=\chi_g(P)$).
This allows \eq{f6}  to be written as 
\be
\int (dU)(UV)_e(WU^\dagger)_e&=&\sum_{P\in S_p}\sum_f{d_f^2\over
p!\b d_f}\chi_f(P)(VW)_P\nn\\
&=&\sum_{R,S'\in S_p}\sum_f{d_f^2\over
(p!)^2\b d_f}\chi_f(RS')(VW)_{RS'}.
\labels{f8}\ee
Introducing the shorthand
\be
\d_{KI_R}\d_{JL_{S'}}=\(\prod_{a=1}^p\d_{k_ai_{R(a)}}\d_{j_a\ell_{S'(a)}}\),\labels{f9a}\ee
the sum over $R$ and $S'$ on the right-hand side is
\be
\sum_{R,S'}\chi_f(RS')V_{LK}W_{IJ}\d_{KI}\d_{JL_{RS'}}&=&
\sum_{R,S'}\chi_f(RS')V_{LK}W_{IJ}\d_{KI}\d_{J_{R^{-1}}L_{S'}}\nn\\
&=&\sum_{R,S'}\chi_f(RS')V_{LK}W_{IJ_R}\d_{KI}\d_{JL_{S'}}\nn\\
&=&\sum_{R,S'}\chi_f(RS')V_{LK}W_{I_{R^{-1}}J}\d_{KI}\d_{JL_{S'}}\nn\\
&=&\sum_{R,S'}\chi_f(RS')V_{LK}W_{IJ}\d_{KI_R}\d_{JL_{S'}}.\labels{f9}\ee

Since $V$ and $W$ are arbitrary, the coefficients of $V_{LK}W_{IJ}$ on both sides must be the same.
Hence
\be
\bk{I,J|K,L}=\int(dU)U^*_{IJ}U_{KL}=\sum_f\sum_{R,S'}{d_f^2\over (p!)^2\b d_f}\d_{KI_R}\d_{JL_{S'}}\chi_f(RS\,').\labels{f10}\ee

Let us now apply \eq{f10} to the special case $\bk{I,J|I,J_Q}$. Since $K=I$, the factor $\d_{I\,I_R}$
is non-zero if and only if $R\in{\cal G}_I$. Similarly, since $L=J_Q$, the factor
$\d_{JL_{S'}}$ equals to $\d_{JJ_{QS'}}$. 
Hence $S=QS'$ must be in the invariant group ${\cal G}_J$.
Now the argument of $\chi_f$ in \eq{f10} is $RS'=RQ^{-1}S$. Since $R$ is summed over
a group and so is $S$, we may replace $R$ by $R^{-1}$ and $S$ by $S^{-1}$. With
this replacement, the argument of $\chi_f$ is $(SQR)^{-1}$, Using $\chi_f(SQR)
=\chi_f((SQR)^{-1})$, we finally arrive at the formula shown in \eq{gpint}.

Formula \eq{gpint} can also be obtained from the 
following version of the Itzykson-Zuber formula
\be
\int (dU) \exp\[\beta \Tr\(M_1UM_2U^\dagger\)\]=\sum_f{\beta^{|f|}d_f\over|f|!\b d_f}
\b\chi_f(M_1)\b\chi_f(M_2),\labels{iz}\ee
where the sum is over all the Young's tableaux, and $|f|=\sum_if_i$ is the number of boxes
in a particular tableau. Using the orthonormal relation \eq{f7},
Frobenius's formula \eq{frobenius} can be inverted to read
\be
\int (dU) \exp\[\beta \Tr\(M_1UM_2U^\dagger\)\]=\sum_f{\beta^{|f|}d_f\over(|f|!)^3\b d_f}
\sum_{R,S}\chi_f(R)\chi_f(S)(M_1)_R(M_2)_S.\labels{f12}\ee
Next, use the formula
\be
\chi_f(R)\chi_f(S)={d_f\over|f|!}\sum_P\chi_f(RPSP^{-1}),\labels{f13}\ee
which can be derived from \eq{f2}, to combine the two characters into one,
\be
\int (dU) \exp\[\beta \Tr\(M_1UM_2U^\dagger\)\]&=&\sum_f{\beta^{|f|}d_f^2\over(|f|!)^4\b d_f}
\sum_{P,R,S}\chi_f(RPSP^{-1})(M_1)_R(M_2)_S\nn\\
&=&\sum_f{\beta^{|f|}d_f^2\over(|f|!)^4\b d_f}
\sum_{P,R,S'}\chi_f(RS')(M_1)_R(M_2)_{P^{-1}S'P}.\labels{f14}\ee
Since $(M_2)_S$ depends only on the class that
 $S$ lies in, the last factor is thus independent of $P$.
Therefore the sum over $P$ yields only a factor $|f|!$.

Identifying terms proportional to $\beta^p$ on both sides, one gets
\be
\sum_{I,J,K,L}\bk{I,J|K,L}(M_1)_{IK}(M_2)_{LJ}&=&
\sum_f{d_f^2\over(p!)^2\b d_f}
\sum_{R,S'}\chi_f(RS')(M_1)_{IK}(M_2)_{LJ}\d_{KI_R}\d_{JL_{S'}},\labels{f15}\ee
which agrees with \eq{f10}.


\newpage
\begin{center}
{\bf References}
\end{center}

\begin{thebibliography}{9}
\bibitem{DGZ-J} P. Di~Francesco, P.~Ginsparg, and J. Zinn-Justin, Phys.~Rep. 254 (1995) 1. 
\bibitem{GM-GW} A. Mueller-Groeling, and H.A. Weidenmueller, Phys.~Rep. 299 (1998) 189 cond-mat/9707301. 
\bibitem{LMZ} G. Mahlon, C.S. Lam, and W. Zhu, Phys.~Rev.~ D66  (2002)  074005.
\bibitem{Cr} M. Creutz, J.~Math.~Phys. 19 (1978) 2043.
\bibitem{IZ} C. Itzykson and J.-B. Zuber, J.~Math.~Phys. 21 (1980) 411.
\bibitem{GT} S. Samuel, J.~Math.~Phys. 21 (1980) 2695;
I. Bars, J.~Math.~Phys. 21 (1980) 2678; I. Bars, Phys.~Scripta 23 (1981) 983;
A.~Morozov Mod.~Phys.~Lett. A7 (1992) 3503; 
A.B. Balantekin, Phys.~Rev. D62 (2000) 085017;
B.  Collins, math-ph/0205010;
B. Schlittgen and T.~Wettig, J.~Phys. A36 (2003) 3195;
P. Zinn-Justin and J.-B.~Zuber, J.~Phys. A36 (2003) 3173.
\bibitem{Weyl} H.~Weyl, `The Classical Groups', (Princeton University Press, 1966).
See theorems (7.5.B) and (7.7.A).
\bibitem{LN}
P. Rossi, M. Campostrini, and E. Vicari, Phys.~Rep. 302 (1998) 143;
A. Matytsin, Nucl.~Phys.~B 411 (1994) 805. 
\end{thebibliography}
\end{document}